\documentclass[aps,prapplied,reprint,groupedaddress]{revtex4-1}
\usepackage{amsmath,amssymb}
\usepackage{bm}
\usepackage{multirow}
\usepackage{xcolor}
\usepackage{srcltx}
\usepackage[normalem]{ulem}
\usepackage{graphicx}
\usepackage[utf8]{inputenc}
\usepackage{epstopdf}
\usepackage[separate-uncertainty=true]{siunitx}

\makeatletter
\let\Hy@backout\@gobble
\makeatother

\begin{document}


\title{High efficiency \lq plug \& play\rq\space source of heralded single photons}


\author{Nicola Montaut}
\email{montaut@mail.upb.de}
\author{Linda Sansoni}
\author{Evan Meyer-Scott}
\author{Raimund Ricken}
\author{Viktor Quiring}
\author{Harald Herrmann}
\author{Christine Silberhorn}


\affiliation{Integrated Quantum Optics, Universit\"at Paderborn, Warburger Strasse 100, 33098 Paderborn, Germany}



\begin{abstract}
Reliable generation of single photons is of key importance for fundamental physical experiments and to demonstrate quantum technologies. Waveguide-based photon pair sources have shown great promise in this regard due to their large degree of spectral tunability, high generation rates and long photon coherence times. However, for such a source to have real world applications it needs to be efficiently integrated with fiber-optic networks. We answer this challenge by presenting an alignment-free source of photon pairs in the telecommunications band that maintains heralding efficiency $> \SI{50}{\percent}$ even after fiber pigtailing, photon separation, and pump suppression.
The source combines this outstanding performance in heralding efficiency with a compact, stable, and easy-to-use \lq plug \& play\rq\space package: one simply connects a laser to the input and detectors to the output and the source is ready to use. This high performance can be achieved even outside the lab without the need for alignment which makes the source extremely useful for any experiment or demonstration needing heralded single photons.
\end{abstract}


\maketitle

\section{Introduction}
Quantum optical technologies have seen a shift in the last decade from table-scale bulk optics to integrated circuits, greatly increasing the capability to implement quantum information tasks with reliable, compact and stable devices.
Many integrated quantum devices have been fabricated, ranging from waveguide sources \cite{tanzilli2001, fujii2007, zhang2007, zaske2011, nosaka2006,ghalbouni2013} and entanglement sources~\cite{koenig2005, jiang2007, arahira2011, ramelow2013, herrmann2013, herbauts2013} to on-chip detectors~\cite{gerrits2011,Ferrari2016}, from quantum teleporters~\cite{metcalf2014} to complex linear circuits~\cite{carolan2015} with photonic manipulation~\cite{jin2014}.
However, the functioning of these circuits still relies on the careful alignment of external bulk components, limiting usefulness outside the lab. 
Here we dispense with lengthy and tedious optical alignments by realizing a \lq plug \& play\rq\space photon source for quantum information applications that maintains laboratory-scale high performance.
Our source of telecom-wavelength heralded single photons based on parametric down-conversion (PDC) requires no user intervention, and produces photons with high heralding efficiency and picosecond coherence time.  
To remove the need of bulk components we adopt a hybrid platform, exploiting the high nonlinearity and low loss of periodically-poled titanium-indiffused lithium niobate (Ti:PPLN) waveguides to produce photon pairs with high brightness, and standard telecommunications components for photon routing and filtering. The platforms are connected with a low-loss fiber pigtail. The lithium niobate and fiber-optic sections thus form a fully integrated source, producing single photons in single-mode fiber already filtered and seperated. It provides the necessary link to bridge the gap between experiments conducted in academic laboratories and real-world quantum applications.

Since quantum states cannot be deterministically amplified~\cite{wooters1982}, reducing losses is much more critical than for classical systems. Thus low-loss interfaces between sources and optical fiber networks are vital for many applications \cite{jiang2007, arahira2011, arahira2012, meany2014, kaiser2016} and permanent fixation of fibers is beneficial for ease of use and stability. However, maintaining high efficiency during fiber packaging has proved to be quite challenging. Previous fiber-pigtailed photon-pair sources~\cite{zhong2009,oesterling2015} showed raw heralding efficiencies $\lesssim \SI{3}{\percent}$, which we improve upon by more than an order of magnitude. In addition, our source generates photons in the picosecond coherence range. This intermediate bandwidth regime shows promise in long distance applications~\cite{sedziak2017} as they exhibit low dispersion when transmitted through fibers and can be filtered down to produce spectrally pure states without compromising on source brightness.

\begin{figure*}
\centering
\includegraphics[width=0.9\textwidth, trim={11.5cm 0 17cm 4cm},clip]{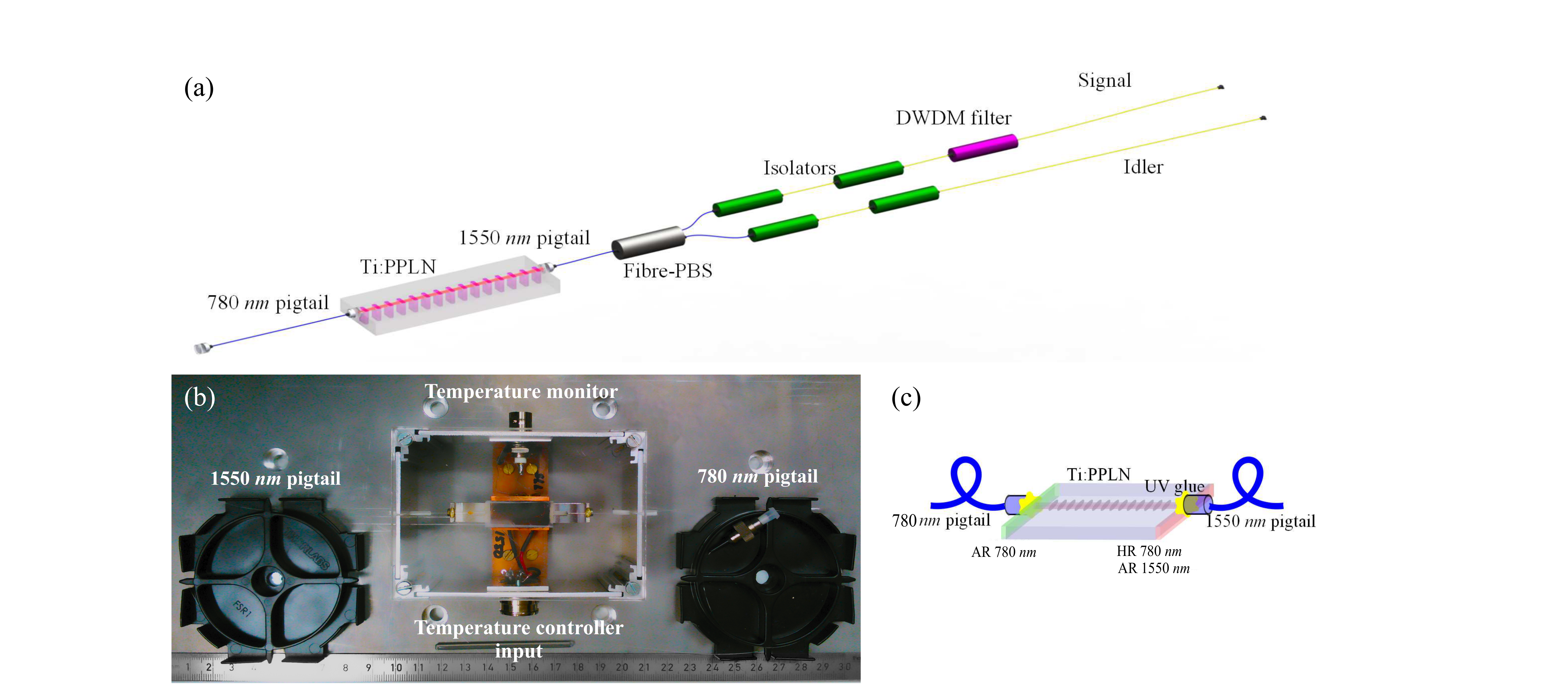}
\caption{(a) Schematic of the \lq plug \& play\rq\space heralded single photon source. (b) Image of the packaged waveguide chip. (c) Depiction of pigtailed chip. Abbreviations: PBS, polarizing beam splitter; DWDM filter, dense wavelength division multiplexing filter; AR, anti-reflection; HR, high reflection}
\label{Fig:Source}
\end{figure*}

\section{Device Design}
Our device takes advantage of high-efficiency parametric down-conversion (PDC) based on the $\chi^{(2)}$ nonlinearity in waveguides. Periodically poled titanium indiffused waveguides in lithium niobate exhibit extremely low losses \cite{schmidt1974,regener1985} and provide optical guiding in TE and TM polarizations, thereby allowing type II PDC wherein an ordinarily polarized pump beam decays to orthogonally polarized photon pairs. This provides easy separation of degenerate PDC photons based on their polarization, unlike type 0 sources~\cite{tanzilli2001, tanzilli2002, hubel2010, oesterling2015} that split photons using 3 dB couplers which results in the loss of half of the generated photons. Additionally type II phasematching provides intrinsically narrowband photons.

The fully fiber-pigtailed device is shown in Fig. \ref{Fig:Source} (a), (b). The Ti:PPLN waveguide chip can be temperature-tuned via a built-in Peltier element. The waveguide is pigtailed with polarization-maintaining fibers (PMFs) on both end facets. The output fiber is spliced to a fiber polarizing beam splitter (fiber-PBS), which separates the signal and idler into two spatial modes. Fiber isolators after the PBS ensure pump suppression and a filter in the signal arm is used to remove noise and increase spectral purity.
\\
\section{Device engineering}
The waveguide and poling were designed to allow phasematched PDC for a pump around \SI{780}{\nano\meter} to degenerate photons around \SI{1560}{\nano\meter}, and to optimize couping to standard single mode fibers. Fabrication proceeded as follows: \SI{25}{\milli\meter} long waveguides were produced by in-diffusing a titanium strip of \SI{7}{\micro\meter} width and \SI{80}{\nano\meter} thickness on a lithium niobate substrate at \SI{1060}{\celsius} for \SI{8.5}{\hour}. Next the waveguides were periodically poled with period \SI{9.08}{\micro\meter} for a length of \SI{21}{\milli\meter}. The input facet of the chip received an anti-reflective coating for the pump, and the output facet received a  high-reflective coating for the pump providing $\approx\SI{22}{\decibel}$ suppression and an anti-reflective coating for the photon pairs, providing maximum coupling to the pigtailed fiber. 

\begin{figure*}
\centering
\includegraphics[width=\textwidth]{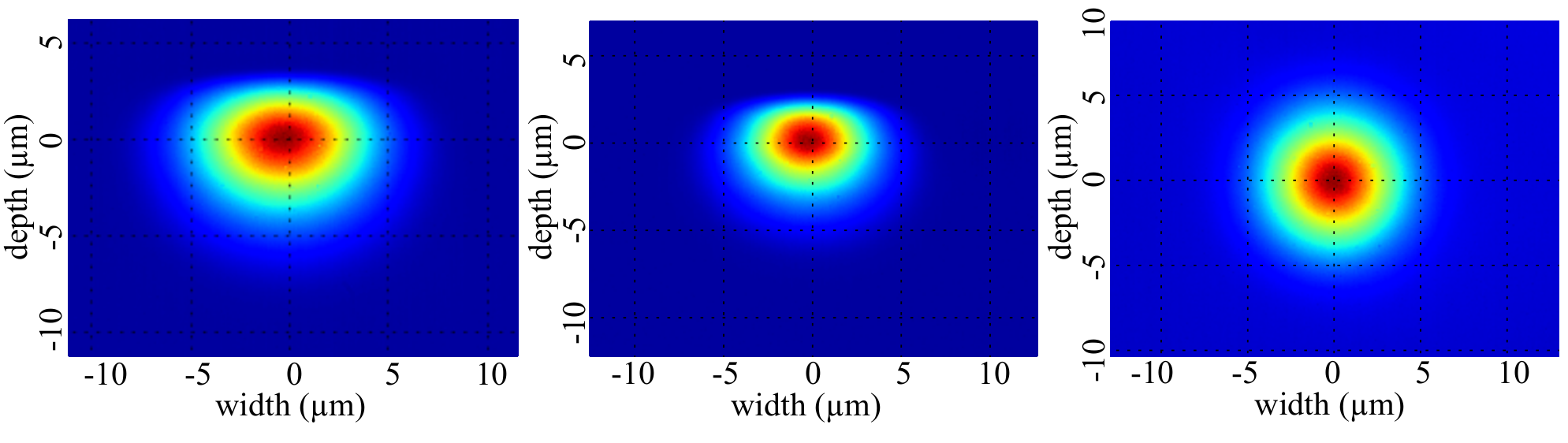}
\caption{Measured optical field profiles of the TE (left), TM (centre) modes in waveguide and PANDA 1550\ nm fiber mode (right).}
\label{Fig:Modes}
\end{figure*}

A crucial step of source preparation was to pigtail the front and rear ends by permanently fixing PMFs to the waveguide. To achieve the maximum coupling efficiency we first measured the mode profiles of the waveguide (shown for both polarizations in Fig.~\ref{Fig:Modes}) and standard \SI{1550}{\nano\meter} PMFs. The width and height of the TE mode of a \SI{7}{\micro\meter} waveguide were measured to be \SI{7.0}{\micro\meter} and \SI{4.7}{\micro\meter}, and for the TM mode \SI{5.3}{\micro\meter} and \SI{3.4}{\micro\meter} respectively. The measured mode diameter of a PMF mode was \SI{6.08}{\micro\meter}. With these data the mode profiles were modelled (FemSIM package, RSoft) to find maximum achievable coupling efficiencies between waveguide and fiber modes for TE and TM of \SI{92}{\percent} and \SI{85}{\percent} respectively.

The pigtailing of the output facet of the chip with a PMF was then carried out as follows: broadband light at telecom wavelength was injected into the waveguide and collected using a PMF. The position of the fiber was controlled using a six-axis motorized stage that is capable of scanning in sub-micron resolutions along the end face of the waveguide. After performing multiple scan iterations of increasing resolution, the optimal position of the fiber was determined when maximum intensity of light coupled from the waveguide to the fiber for both TE and TM polarizations which was detected using a fiber-PBS and InGaAs photodiodes. Once the highest coupling was achieved, ultraviolet curing glue (Norland Products Inc., NOA-81) was applied at the interface between the PMF and chip and cured with an ultraviolet source (see Fig. \ref{Fig:Source} (c)). During the curing process, the position of the fiber shifts slightly thereby resulting in a decreased pigtailing efficiency of \SI{84.0}{\percent} for TE and \SI{75.7}{\percent} for TM. The pigtailed chip is packaged and temperature stabilized via a thermoelectric cooler (TEC) which operates at temperatures up to~\SI{70}{\celsius}.
The input facet of the waveguide was then pigtailed to a PMF for \SI{780}{\nano\meter} with the same positioning apparatus by exploiting second harmonic generation (SHG) to the pump wavelength ensure maximum coupling into the fundamental pump mode of the waveguide. The coupling efficiency between fiber and waveguide for pump wavelengths was measured to be  $\eta_{pump}>\SI{30}{\percent}$.

In order to separate the generated twin photons, we spliced the output PMF to a fiber polarization beam splitter. A pair of C-band in-line isolators were then spliced to each arm of the fiber-PBS to ensure pump suppression: each isolator provides \SIrange{40}{60}{\decibel} suppression for the pump due to the presence of a crystal that absorbs the pump. A fiber dense-wave-division multiplexing (DWDM) filter with a bandwidth of \SI{200}{\giga\hertz} ({\SI{1.6}{\nano\meter}}) was also attached to the signal arm to remove noise and improve the purity of the heralding photons. To ensure maximum transmission the idler arm remained unfiltered.
The final device is completely alignment free and operates as a \lq plug \& play\rq\space source of heralded single photons.
\\
\section{Results and Discussion}
\subsection{Classical characterization}

\begin{figure}[h]
\centering
\includegraphics[width=1.05\columnwidth,trim={0 7.2cm 0 7.2cm},clip, height=7cm]{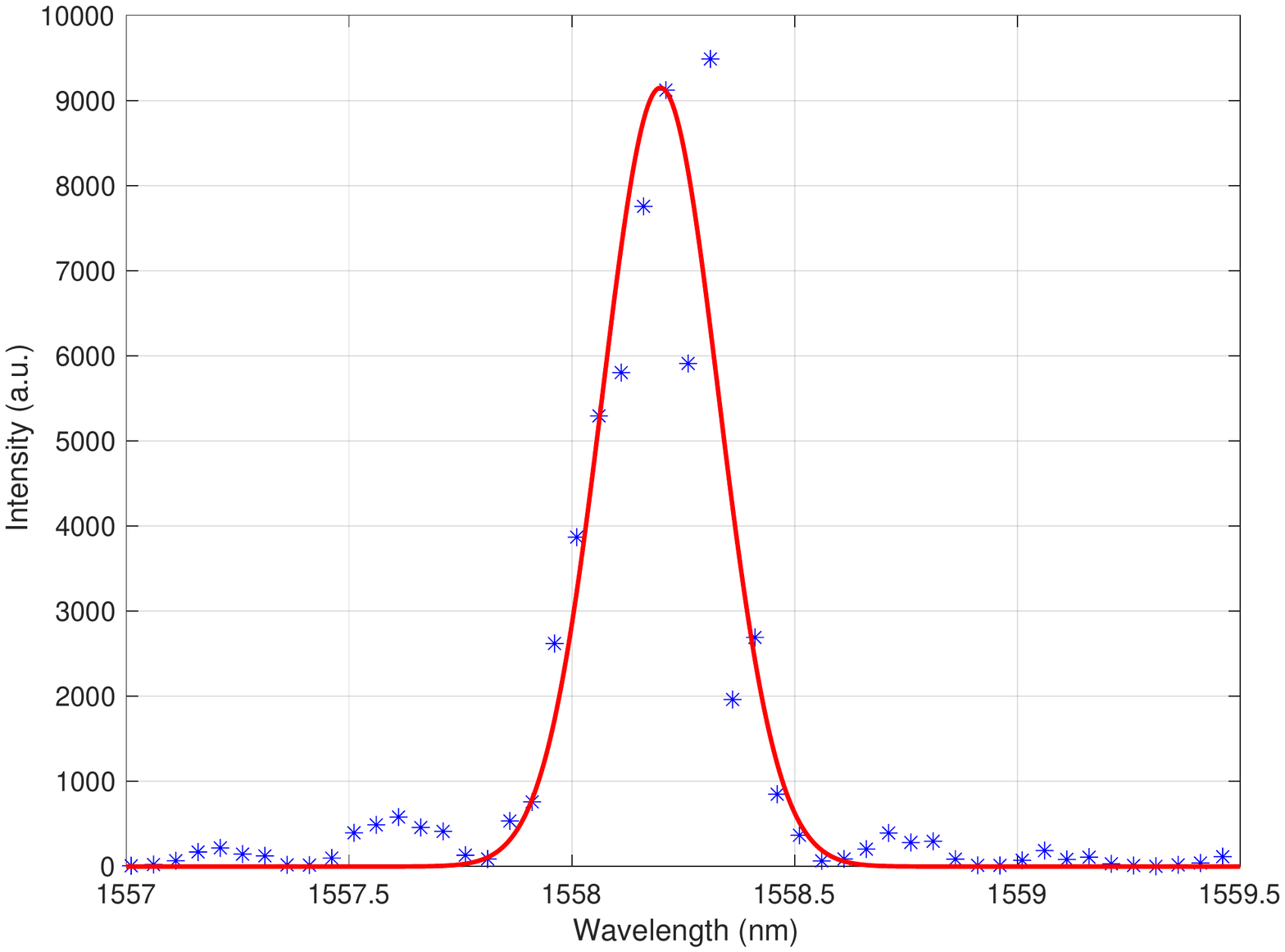}
\caption{Second harmonic generation intensity of the Ti:PPLN waveguide before pigtailing, plotted versus SHG pump wavelength.}
\label{Fig:SHG}
\end{figure}

We first measured the nonlinear behaviour and the losses of the source classically.
To determine the nonlinear conversion efficiency and the spectral degeneracy point we used second harmonic generation (SHG) from the telecom to pump wavelength. A tunable continuous-wave laser at telecom wavelength was injected at \SI{45}{\degree} polarization into the waveguide and the generated light at twice the frequency was detected using a Si PIN-diode. Phasematching was observed at \SI{1558.29}{\nano\meter}, as shown in Fig. \ref{Fig:SHG} at a temperature of \SI{25}{\celsius} with a full-width at half-maximum (FWHM) bandwidth of \SI{0.31\pm0.02}{\nano\meter} and an efficiency of \SI{2.56\pm0.23}{\percent\per\watt\per\centi\meter\squared}. We thus adopted this phasematching condition for the production of heralded single photons.

We next measured the losses in the photon path classically, to compare with the obtained heralding efficiencies (below).
The measured waveguide loss \cite{regener1985} at \SI{1550}{\nano\meter} for TE was \SI{0.13\pm0.02}{\decibel/\centi\meter} and for TM was \SI{0.18\pm0.02}{\decibel/\centi\meter}. The fiber-PBS was spliced to a pair of isolators on each arm and the measured transmissions of the PBS-isolators combination were \SI{84.0\pm1.9}{\percent} and \SI{81.5\pm1.9}{\percent} respectively. A fiber-based DWDM filter with a transmission of \SI{86.0\pm2.0}{\percent} attached to the signal arm was used as the heralding arm for the PDC process. Therefore, the maximum heralding efficiency achievable in the idler arm was calculated to be {$\eta_i^{calc}= \mathrm{T_{wg,TM} \cdot T_{FP,TM} \cdot T_{PBS}} = \SI{58.6\pm1.4}{\percent}$} before detection, where $\mathrm{T_{wg,TM}}$, $\mathrm{ T_{FP,TM}}$ and $\mathrm{T_{PBS}}$ is the transmission of the waveguide, fiber pigtail and the PBS-isolator unit respectively for TM polarization.

\begin{figure*}
\centering
\includegraphics[width=0.9\textwidth,trim={0 13cm 0 1.5cm},clip, height=11cm]{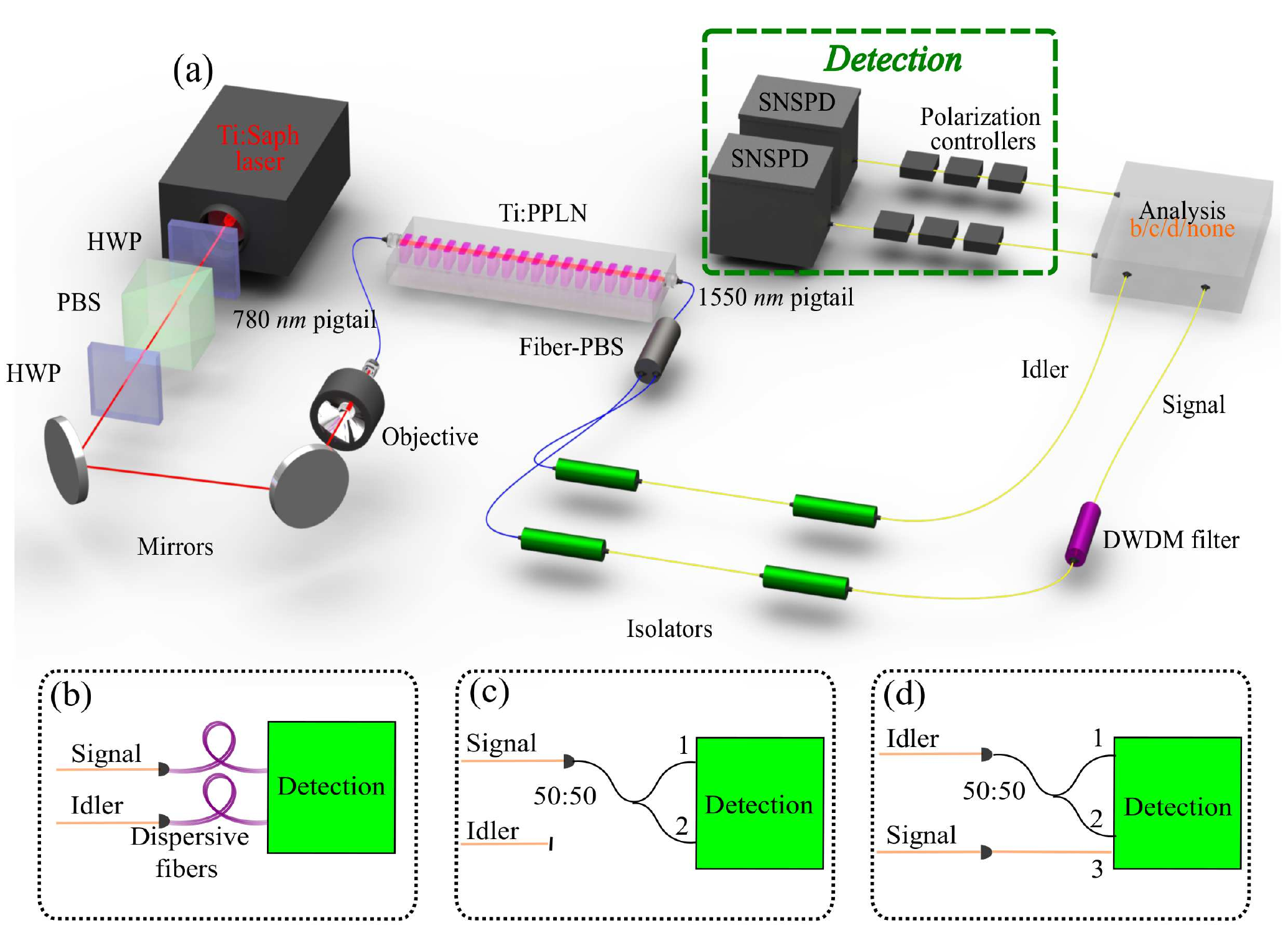}
\caption{(a) Schematic of the experimental setup used for efficiency measurements: The generated signal and idler photons pass directly to the detection unit or through one of the four quantum setups. (b) JSI measurement. (c) Unheralded $g^{(2)}(0)$ measurement. (d) Heralded $g^{(2)}(0)$ measurement.} 
\label{Fig:Setup}
\end{figure*}
\begin{table*}[t]
\centering
\begin{tabular}{|p{2.3cm}|p{3.5cm}p{2.5cm}rrrrrrrrrrrrrrrrrr|}
 \hline
 &Reference & Source & $\lambda$ &&&&&$\Delta\lambda$&&&&& $\eta_{s/i}$&&&&& $\eta_{s/i}^{{corr}}$&&\\
 &&&($nm$)&&&&& ($nm$)&&&&&(\%)&&&&&(\%)&&
 \\
 \hline
 \hline
 \multirow{7}{*}{FREESPACE}&
 Soujaeff et al~\cite{soujaeff2007}& BBO/bulk&1550&&&&&18 &&&&&$1.9^*$&&&&&18.7&&\\ 
 &Kurtsiefer et al~\cite{kurtsiefer2001}& BBO/bulk &700 &&&&& 4.6 &&&&&$6.1^* $&&&&&28.6&&\\
 &Alibart et al~\cite{alibart2005} &PPLN/wg &1550 &&&&& 20&&&&& $3.7^*$ &&&&&37&&\\
 &Pittmann et al~\cite{pittman2005}& BBO/bulk &780 &&&&&9&&&&&31&&&&& $49.2^*$&&\\
 &Giustina et al~\cite{giustina2015} &PPKTP/bulk &810&&&&&- &&&&&78.6 &&&&&$82.7^*$&&\\
 &Shalm et al~\cite{shalm2015}&PPKTP/bulk&1550&&&&&-&&&&&75.6&&&&&$83.1^*$&&\\
 &U'Ren et al~\cite{uren2004} & PPKTP/wg&800&&&&&2 &&&&&51 &&&&& 85&&\\
 \hline
 \hline
 \multirow{3}{*}{PIGTAILED}&
 Zhong et al~\cite{zhong2009}& PPKTP/wg &1316&&&&& 1.3 &&&&&2.8&&&&& $13.7^*$&&\\
 &Oesterling et al~\cite{oesterling2015}&PPLN/wg&1550&&&&& 60 &&&&&$3.1^*$&&&&& $30.9^*$&&\\
 &\textbf{This work} & \textbf{PPLN/wg}  &\textbf{1560} &&&&& \textbf{1.8} &&&&& \textbf{46.2}&&&&& \textbf{54.4}&&\\
 \hline 
\end{tabular}
\caption{Comparison of different sources with respect to their photon bandwidth ($\Delta\lambda$), raw and detector-corrected heralding efficiencies. $^*$ has been estimated from reported data.}
\label{Tab:sources}
\end{table*}

\subsection{Quantum characterization}
We benchmark here the key performance metrics of our source: heralding efficiency, brightness, spectral purity, and single-photon coherence time, showing they reach or surpass the state-of-the-art even after full fiber integration. The setup used for the quantum measurements is shown in Fig. \ref{Fig:Setup} (a). A mode-locked Ti:Sapphire laser that generates picosecond pulses with FWHM bandwidth \SI{0.3}{\nano\meter} at \SI{779.15}{\nano\meter} wavelength and \SI{1}{\mega\hertz} repetition rate was used as the pump, with coupled powers of \SIrange{1}{30}{\micro\watt}. Photons were detected with superconducting nanowire single photon detectors (SNSPDs, PhotonSpot) with detection efficiencies of $\eta_{det}=\SI{85}{\percent}$.

\begin{figure*}
\centering
\includegraphics[width=0.9\textwidth, height=11cm]{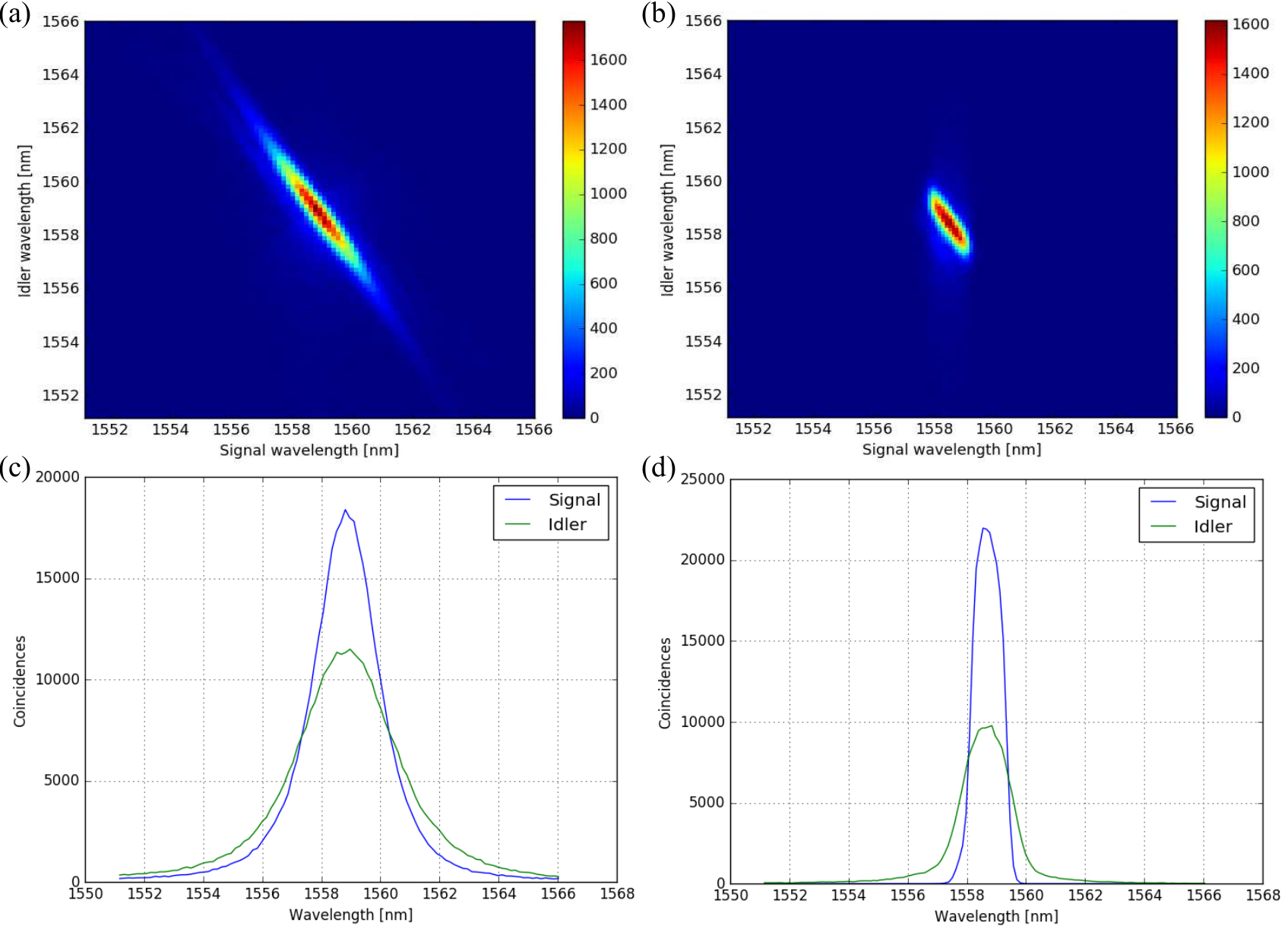}
\caption{Joint spectral intensities and marginals of (a) \& (c) unfiltered PDC, (b) \& (d) filter attached only on the signal arm. The marginals are the projection directly from the JSI, and thus represent the heralded marginal spectra.}
\label{Fig:JSI}
\end{figure*} 
The heralding efficiency of the source was determined using the apparatus as shown in Fig. \ref{Fig:Setup} (a) where the outputs of the device are directly connected to the SNSPDs. The raw heralding efficiency of the idler photon is given by $\eta_{i}=\frac{C}{S_{s}}$, where $C$ are coincidences between the two outputs, $S$ are single counts, and $s,i$ label signal and idler, respectively. This was measured to be $\eta_i=\SI{46.2\pm0.6}{\percent}$. On correcting the raw value for detector efficiency, but retaining the channel losses (in contrast with~\cite{oesterling2015}) we estimate the efficiency of the complete device from generation to output to be $\eta_i^{corr}=\frac{\eta_i}{\eta_{det}}=\SI{54.4\pm0.7}{\percent}$, which is very close to the calculated maximum heralding efficiency that can be achieved from this device. In table \ref{Tab:sources} we compare our source with other single photon sources. Our results represent the best raw and detector-corrected heralding efficiencies when compared to other fiber pigtailed sources~\cite{zhong2009,oesterling2015} and closely approach the quality of state-of-the-art bulk sources. However there is still room for improvement, particularly in the PBS and filters. Here further integration would help, in particular integrated PBSs \cite{sansoni2016} and/or wavelength division multiplexers~\cite{zhang2007,jin2014} on chip for splitting of the photon pairs and pump suppression respectively.

The arrangement used to measure the joint spectral intensity (JSI) of the PDC photons with time-of-flight spectrometers~\cite{avenhaus2009} (Fig. \ref{Fig:Setup} (b)). The JSI and marginals of filtered and unfiltered PDC photons are shown in Fig. \ref{Fig:JSI}. The spectral bandwidth of the signal and idler photons were found to be $\Delta\lambda_s=\SI{2.80\pm0.12}{\nano\meter}$ and $\Delta\lambda_i=\SI{3.68\pm0.27}{\nano\meter}$ respectively (See Fig. \ref{Fig:JSI} (a), (c)). On applying the filter to the signal arm, the measured FWHM bandwidth was $\Delta\lambda_s=\SI{1.06\pm0.02}{\nano\meter}$ and $\Delta\lambda_i=\SI{1.83\pm0.04}{\nano\meter}$ for signal and idler arms respectively (See Fig. \ref{Fig:JSI} (b), (d)). The photon bandwidth of some other sources are listed in table \ref{Tab:sources}. By using the spectrum from the JSI measurement, we determined the single-photon coherence time of the idler photon by performing a Fourier-transform on the idler marginals on assuming a delta function transmission of the signal filter. The single-photon coherence time was estimated to be $\Delta\tau_{sp} = \SI{5.8\pm0.3}{\pico\second}$. From this we calculated that the generated single photon can travel $> \SI{130}{\kilo\meter}$ in a standard optical fiber before it starts to overlap with its neighbouring pulses for a $\SI{1}{\giga\hertz}$ repetition rate. This is nearly an order of magnitude higher than equivalent possible distance for femtosecond photons~\cite{ansari2014}, a gap maintained independent of repetition rates. This makes picosecond sources such as ours strongly desirable for long distance quantum communication applications.

The brightness of the chip was measured to be $\mathcal{B}_{chip}= \frac{S_s \cdot S_i}{C \cdot t \cdot P_{chip} \cdot \Delta\lambda_s} = (1.39\pm0.04)\cdot10^7 \mathrm{\frac{\ pairs}{\ s\ \cdot\ mW \cdot\ nm}}$ at a mean photon number $\langle n\rangle \approx 0.01\ \mathrm{\frac{photons}{pulse}}$, where $t$ is the integration time, $P_{chip}$ is the pump power coupled into the chip and $\Delta\lambda_s$ is the bandwidth of the signal photon obtained from the JSI. The source shows higher or similar photon generation rates when compared with other waveguide SPDC sources~\cite{zhong2009, harder2013, herrmann2013, oesterling2015, kaiser2016, sansoni2016}. 

The spectral purity of the single photon source can be estimated from the joint spectra \cite{cassemiro2010} which was determined to be $\mathcal{P}=0.69$. We also directly characterized the purity of the source by the unheralded second order Glauber correlation function, $g^{(2)}(0)$. If the JSI is highly correlated, that is consisting of multiple frequency modes, we expect a Poissonian photon number distribution, corresponding to $g^{(2)}_P(0)=1$. Alternatively, a de-correlated PDC spectrum, corresponding to a spectrally pure state, results in a thermal distribution with $g^{(2)}_{th}(0)=2$. Since our photons are in a single spatial mode, they will have the same spectral purity in coincidence, allowing us to measure the signal's purity after filtering to determine the idler's purity after heralding. The setup required to measure the $g^{(2)}(0)$ is as shown in Fig. \ref{Fig:Setup} (c), where the signal photons are split by a 50:50 coupler then detected, giving $g^{(2)}_{s,raw}(0) = 1.37\pm0.01$. Noise counts arising from fluorescence or other parasitic processes present in the signal strongly degrade the $g^{(2)}(0)$ value. The background events in both arms of the splitter were measured to be $\SI{9.13\pm0.07}{\percent}$ of the singles counts. On correcting for these background counts~\cite{eckstein2011g2}, we get $g^{(2)}_{s}(0) = 1.66\pm0.05$ corresponding to a Schmidt number \cite{eckstein2011, christ2011} $\mathcal{K}=1.52$ and a purity, $\mathcal{P}=0.66\pm0.05$, which matches closely with the estimation from the JSI. The unfiltered idler arm showed a $g^{(2)}(0)$ of $g^{(2)}_{i}(0) = 1.07\pm0.03$ as expected from the highly correlated spectrum (see Fig. \ref{Fig:JSI} (a)). The $g^{(2)}(0)$ can be improved by using filters with smaller bandwidth to minimize the correlations still present in the filtered spectrum; however, this comes at the cost of lowering the brightness and heralding efficiency of the source~\cite{meyer2017}.

To quantify the contributions from higher order photon emissions from the source, a heralded $g^{(2)}(0)$ measurement was carried out. An ideal single photon source gives heralded $g^{(2)}_{h}(0)$ = 0. Now, the other photon is also detected and the coincidences between all three detection events are analysed (Fig. \ref{Fig:Setup} (d)). The heralded second correlation was determined by $g^{(2)}_{h}(0) = \frac{C_{1,2,3}*N_{3}}{C_{1,3}*C_{2,3}}$, where the number of two- and three-fold coincidences are given by $C_{l,k}$ and $C_{l,k,m}$ respectively, where $l,k,m=\left\{1,2,3\right\}$ label the 3 outputs and $N_3$ denotes the number of events in the heralding (signal) arm. We measured $g^{(2)}_{h,i}(0)= 0.014\pm0.001$ at $\langle n\rangle \approx 0.002\ \mathrm{\frac{photons}{pulse}}$ for the idler, well below the threshold commonly used for single photons ($g^{(2)}_h(0)=0.5$ \cite{leifgen2014}).
\\
\section{Conclusion}
We have experimentally demonstrated an efficient picosecond heralded single photon source via type II PDC in Ti:PPLN waveguides that is fully fiber integrated (including photon separation, pump suppression and filtering), packaged and ready to integrate with complex quantum systems. This device presents great advantages in terms of ease of use and stable operation, together with quantum performances which exceed by far the packaged devices available up to now: high brightness, low losses, high heralding efficiencies and the ability to produce picosecond photons in the telecom range. Owing to mature source engineering technology, the device offers complete flexibility to produce a large range of PDC wavelengths and operation at non-degeneracy, and can be integrated with a variety of filters for user-specific applications. In order to achieve a completely integrated device, the source can also be implemented by pumping with commercially available semiconductor lasers instead of a pulsed system as shown in this work.
The quantum characteristics combined with the \lq plug \& play\rq\space configuration of this source represent a great achievement in the development of quantum technology for daily-use applications.

\section{acknowledgments}
The authors would like to thank Thomas Nitsche, Georg Harder and Kai Hong Luo for their helpful comments and technical support. This work was supported by the Marie Curie Initial Training Network PICQUE (Photonic Integrated Compound Quantum Encoding, grant agreement no. 608062, funding Program: FP7-PEOPLE-2013-ITN, http://www.picque.eu), DFG (Deutsche Forschungs-gemeinschaft, grant no. SI 1115/4-1 and Gottfried Wilhelm Leibniz-Preis) and NSERC (The Natural Sciences and Engineering Research Council of Canada).

%

\end{document}